\documentclass{article}

\usepackage{emulateapj,apjfonts}

\input epsf

\lefthead{Janka et al.}
\righthead{BLACK HOLE -- NEUTRON STAR MERGERS}

\def\ave#1{\langle #1 \rangle}

\begin{document}

\title{BLACK HOLE -- NEUTRON STAR MERGERS AS CENTRAL ENGINES OF GAMMA-RAY BURSTS}

\author{H.-Thomas Janka\altaffilmark{1},
	Thomas Eberl\altaffilmark{1,2},
        Maximilian Ruffert\altaffilmark{3},
	and Chris L. Fryer\altaffilmark{4} }

\altaffiltext{1}{Max-Planck-Institut f\"ur Astrophysik, Postfach 1523, D-85748 Garching,
                 Germany; thj@mpa-garching.mpg.de}
\altaffiltext{2}{Technische Universit\"at M\"unchen, Physik-Department E12, 
                 James-Franck-Strasse, D-85748 Garching, Germany; thomas.eberl@physik.tu-muenchen.de}
\altaffiltext{3}{Department of Mathematics and Statistics, University of Edinburgh,
                 Scotland, EH9 3JZ, U.K.; m.ruffert@ed.ac.uk}
\altaffiltext{4}{UCO/Lick Observatory, University of California, Santa Cruz,
                 CA 95064, U.S.A.; cfryer@ucolick.org}

\begin{abstract}
Hydrodynamic simulations of the merger of stellar mass black hole -- neutron
star binaries (BH/NS) are compared with mergers
of binary neutron stars (NS/NS). The simulations are Newtonian,
but take into account the emission and backreaction of gravitational waves.
The use of a physical nuclear equation of
state allows us to include the effects of neutrino emission.
For low neutron star to black hole mass ratios the 
neutron star transfers mass to the black hole during a few cycles of orbital decay
and subsequent widening before finally being disrupted, whereas for ratios
near unity the neutron star is already distroyed during its first approach.
A gas mass between $\sim 0.3\,M_{\odot}$ and $\sim 0.7\,M_{\odot}$ is left
in an accretion torus around the black hole and radiates neutrinos at a
luminosity of several $10^{53}\,$erg/s during an estimated accretion time scale of 
about 0.1$\,$s. The emitted neutrinos and antineutrinos annihilate into $e^\pm$
pairs with efficiencies of 1--3\% percent and rates of up to
$\sim 2\times 10^{52}\,$erg/s, thus depositing an energy 
$E_{\nu\bar\nu} \la 10^{51}\,$erg above the poles of the black hole in a region 
which contains less than $10^{-5}\,M_{\odot}$ of baryonic matter. This could
allow for relativistic expansion with Lorentz factors around 100 and is sufficient
to explain apparent burst
luminosities $L_{\gamma}\sim E_{\nu\bar\nu}/(f_{\Omega}t_{\gamma})$ up to
several $10^{53}\,{\rm erg\,s}^{-1}$ for burst durations
$t_{\gamma}\approx 0.1$--1$\,$s, if the $\gamma$ emission is
collimated in two moderately focussed jets in a fraction
$f_{\Omega} = 2\delta\Omega/(4\pi)\approx 1/100$---$1/10$ of the sky.
\end{abstract}


\section{INTRODUCTION}

BH/NS and NS/NS mergers are discussed as promising candidates for the
origin of gamma-ray bursts (GRBs)
(e.g., Blinnikov et al. 1984; Eichler et al. 1989; Paczy\'nski 1991;
Narayan, Piran, \& Shemi 1992; M\'es\-z\'aros 1999; Fryer, Woosley, \& Hartmann 1999;
Bethe \& Brown 1998, 1999), at least for the subclass of less
complex and less energetic short and hard bursts (Mao, Narayan, \& Piran 1994)
with durations of fractions of a second (Popham, Woosley, \& Fryer 1999;
Ruffert \& Janka 1999). Optical counterparts and afterglows of this subclass 
have not yet been observed. Due to the
presence of a region of very low baryon density above the poles of the
black hole, BH/NS mergers are considered as more favorable sources
than NS/NS mergers (e.g., Portegies Zwart 1998; Brown et al. 1999).

Previous Newtonian SPH simulations of BH/NS mergers using
a polytropic equation of state indicate that the neutron star may slowly lose
gas in many mass transfer cycles 
(Klu\'zniak \& Lee 1998; Lee \& Klu\'zniak 1998,1999). Whether dynamical
instability sets in at a minimum separation (Rasio \& Shapiro 1994;
Lai, Rasio, \& Shapiro 1994) or whether stable Roche lobe overflow takes place, 
however, can depend on the neutron star to black hole mass ratio 
(Bildsten \& Cutler 1992) and the 
properties of the nuclear equation of state, expressed by the adiabatic index
(Ury\=u \& Eriguchi 1999). 

In this {\it Letter} we report about the first Newtonian BH/NS merger simulations 
(Eberl 1998) which were done with a realistic nuclear equation of state
(Lattimer \& Swesty 1991) and which therefore yield information about the 
thermodynamic evolution and the neutrino emission. 
They allow one to compare the strength of the gravitational wave (GW) emission
relative to NS/NS mergers and to investigate neutrino-antineutrino ($\nu\bar\nu$)
annihilation as potential source of energy for GRBs.

\section{NUMERICAL METHODS}

The three-dimensional hydrodynamic simulations were performed with a Eulerian
PPM code using four levels of nested cartesian grids which ensure a good
resolution near the center of mass and a large computational volume simultaneously.
Each grid had 64$^3$ zones, the size of the smallest zone was 0.64 or 0.78$\,$km 
in case of NS/NS and 1.25 or 1.5$\,$km for BH/NS mergers. The zone sizes of the
next coarser grid levels were doubled to cover a volume of 328 or 400$\,$km side length
for NS/NS and 640 or 768$\,$km for BH/NS simulations. GW emission and 
its backreaction on the hydrodynamics were taken into account by the 
method of Blanchet, Damour, \& Sch\"afer (1990) (see Ruffert, Janka, \& Sch\"afer
1996). The neutrino emission and corresponding energy and lepton number changes
of the matter were calculated with an elaborate neutrino leakage scheme 
(Ruffert, Janka, \& Sch\"afer 1996), and $\nu\bar\nu$ annihilation around the
merger was evaluated in a post-processing step (Ruffert et al. 1997).

\section{SIMULATIONS}

Table~1 gives a list of computed NS/NS and BH/NS merger models. Besides the baryonic
mass of the neutron star and the mass of the black hole, the spins of the neutron 
stars were varied. ``Solid'' means synchronously rotating stars, ``none'' irrotational
cases and ``anti'' counter-rotation, i.e., spin vectors opposite to the
direction of the orbital angular momentum. The cool neutron stars have a radius
of about 15$\,$km (Ruffert, Janka, \& Sch\"afer 1996)
and the runs were started with a center-to-center distance of 
42--46$\,$km for NS/NS and with 47$\,$km in case of BH/NS for 
$M_{\rm BH} = 2.5\,M_{\odot}$, 57$\,$km for $M_{\rm BH} = 5\,M_{\odot}$ and 
72$\,$km for $M_{\rm BH} = 10\,M_{\odot}$. The simulations were stopped at a time
$t_{\rm sim}$ between 10$\,$ms and $20\,$ms. The black hole was treated as a 
point mass at the center of a sphere with radius $R_{\rm s} = 2GM_{\rm BH}/c^2$ 
which gas could enter unhindered. Its mass and momentum were updated
along with the accretion of matter. Model TN10, which is added for comparison, 
is a continuation of the NS/NS merger model B64 where at time
$t_{\rm sim} = 10\,$ms the formation of a black hole was assumed and the accretion was
followed for another 5$\,$ms until a steady state was reached
(Ruffert \& Janka 1999).

\section{RESULTS}
 
\subsection{Evolution of BH/NS mergers}

Due to the emission of GWs the orbital separation decreases. 
During its first approach, the neutron star transfers matter to the black
hole at huge rates of several 100 up to $\sim 1000\,M_{\odot}$/s. Within 2--3$\,$ms
it loses 50--75\% of its initial mass. In case of the 2.5$\,M_{\odot}$ black
hole the evolution is catastrophic and the neutron star is 
immediately disrupted (Lattimer \& Schramm 1974).
A mass of 0.2--0.3$\,M_{\odot}$ remains in a thick disk around the black hole
($M_{\rm d}$ in Table~2). In contrast, the orbital distance
increases again for $M_{\rm BH} = 5\,M_{\odot}$ and $M_{\rm BH} = 10\,M_{\odot}$ 
and a significantly less massive neutron star begins a second approach. Again, the
black hole swallows gas at rates of more than $100\,M_{\odot}$/s. Even a third cycle
is possible (Fig.~1). Finally, at a distance $d_{\rm ns}$ and time $t_{\rm ns}$ 
the neutron star with a mass of $M_{\rm ns}^{\rm min}$ is destroyed 
and most of its mass ends up in an accretion disk (Table~2). (In case of NS/NS
mergers $t_{\rm ns}$ means the time when the two density maxima of the stars are
one stellar radius, i.e., $d_{\rm ns} = 15\,$km, apart).
 
The increase of the orbital separation is connected with a strong rise of the
specific (orbital) angular momentum of the gas (Fig.~1). Partly this is due to
the fact that the black hole can capture gas with low specific angular momentum 
first, but mainly because only a fraction of the orbital angular momentum of
the accreted gas is fed into spinning up the black hole. This fraction, which is 
lost for the orbital motion, is proportional to the quantity $\alpha$ in Fig.~2.
Figure~2 is based on the parameterized analysis of
non-conservative mass-transfer by Podsiadlowski, Joss, \& Hsu (1992)
(see also Fryer et al. 1999) assuming that mass ejection from the system is
negligible. It shows that disregarding GW emission, the orbital separation can 
increase for small initial black hole mass only after the neutron star has lost
much mass, while for larger initial $M_{\rm BH}$ and smaller $\alpha$ orbital
widening is easier. Without GWs the separation increases when 
$\alpha < (M_{\rm BH}-M_{\rm NS})/(M_{\rm BH}+M_{\rm NS})$. Including angular
momentum loss by GWs in the point-mass approximation and using the mass-loss
rates from the hydrodynamic models (dashed lines in Fig.~2) yields a
qualitative understanding of the behavior visible in Fig.~1 and suggests
that $\alpha$ is between 0.2 and 0.5.

During the merging a gas mass $\Delta M_{\rm ej}$ of $\sim 10^{-4}\,M_{\odot}$
(in case of counter-rotation and $M_{\rm BH} = 2.5\,M_{\odot}$) to
$\sim 0.1\,M_{\odot}$ (corotation and
$M_{\rm BH} = 10\,M_{\odot}$) is dynamically ejected (Table~2). In the latter 
case the associated angular momentum loss is about 7\%, in all other cases
it is less than 5\% of the total initial angular momentum of the system.
Another fraction of up to 24\% of the initial angular momentum is carried
away by GWs. In Table~2 the rotation parameter $a = Jc/(GM^2)$ is given for the
initial state of the binary system ($a_{\rm i}$) and at the end
of the simulation ($a_{\rm f}$) for the remnant of NS/NS mergers or
for the black hole in BH/NS systems, respectively, provided the black hole did not
have any initial spin. When the whole disk mass $M_{\rm d}$ has been swallowed
by the Kerr black hole, a final value $a_{\rm BH}^\infty$ (Table~2)
will be reached in case of the accretion of a corotating, thin disk with 
maximum radiation efficiency.

The phase of largest mass flow rate to the black hole (between 2 and 5$\,$ms
after the start of the simulations) is connected with a maximum of the GW
luminosity $L_{\rm GW}$ which reaches up to $7\times 10^{55}\,$erg/s (Table~1). 
The peak values of $L_{\rm GW}$ and the wave amplitude $rh$ (for distance
$r$ from the source) increase with the black hole mass. The total energy
$E_{\rm GW}$ radiated in GWs can be as much as $0.1\,M_{\odot}c^2$
for $M_{\rm BH} = 10\,M_{\odot}$.

\subsection{Neutrino Emission and GRBs}

Compressional heating, shear due to numerical viscosity, and dissipation in shocks
heat the gas during accretion to maximum temperatures $kT^{\rm max}$ of several 
10$\,$MeV. Average temperatures are  between 5 and 20$\,$MeV, the higher values
for the less massive and more compact black holes. At these temperatures and at 
densities of 
$10^{10}$--$10^{12}\,$g/cm$^3$ in the accretion flow, electrons are non-degenerate
and positrons abundant. Electron neutrinos and antineutrinos are therefore copiously
created via reactions $p + e^- \to n + \nu_e$ and $n + e^+ \to p + \bar\nu_e$
and dominate the neutrino energy loss from the accreted matter. Dense and hot 
neutron matter is not completely transparent to neutrinos. By taking into 
account the finite diffusion time, the neutrino trapping scheme limits the loss 
of energy and lepton number.

In Table~1 maximum and average values of the luminosities ($L_{\nu_i}^{\rm max}$ 
and $L_{\nu_i}^{\rm av}$, respectively, the latter in brackets)
in the simulated time intervals are listed for $\nu_e$ and $\bar\nu_e$ and 
for the sum of all heavy-lepton neutrinos. The latter are denoted by
$\nu_x \equiv \nu_\mu,\,\bar\nu_\mu,\,\nu_\tau,\,\bar\nu_\tau$ and are
mainly produced by $e^+e^-$ annihilation. 
The total neutrino luminosities $L_\nu(t)$ (Fig.~3)
fluctuate strongly with the varying mass transfer rate to the black hole during the
cycles of orbital decay and widening (compare with Fig.~1).
The total energy $E_\nu$ radiated in neutrinos
in 10--20$\,$ms is typically several $10^{51}\,$erg. Time averages of the
mean energies $\ave{\epsilon}$ of the emitted neutrinos are 
$\sim 15\,$MeV for $\nu_e$, 20$\,$MeV for
$\bar\nu_e$, and 30$\,$MeV for $\nu_x$. Luminosities as well as mean energies,
in particular for smaller black holes, are significantly higher than
in case of NS/NS mergers.

At the end of the simulations, several of the BH/NS models have reached a steady
state, characterized by only a slow growth of the black hole mass with
a nearly constant accretion rate. Corresponding rates $\dot M_{\rm d}$ are 
given in Table~2 and are several $M_\odot$/s. From these we estimate torus life 
times $t_{\rm acc} = M_{\rm d}/\dot M_{\rm d}$ of 50--150$\,$ms. 
Values with $>$ and $<$ signs indicate
cases where the evolution and emission are still strongly time-dependent at
$t_{\rm sim}$. In these cases the accretion torus around the black
hole has also not yet developed axial symmetry. In
all other cases the effective disk viscosity parameter $\alpha_{\rm eff}\sim
v_{\rm r}/v_{\rm Kepler} \sim 3\sqrt{6}R_{\rm s}/(t_{\rm acc}c)$,
evaluated at a representative disk radius of $3R_{\rm s} = 6GM_{\rm BH}/c^2$,
has the same value, 4--$5\times 10^{-3}$. This value is associated with
the numerical viscosity of the hydro code (which solves the Euler equations)
for the chosen resolution. The further disk evolution is driven by the angular
momentum transport mediated by viscous shear forces, which determines the 
accretion rate. 
The physical value of the disk viscosity is unknown. The numerical viscosity of
our code, however, is in the range where the viscous energy dissipation and 
the energy emission by neutrinos should be roughly equal, i.e., where the 
conversion efficiency $q_\nu = \ave{L_\nu}/(\dot M_{\rm d}c^2)$ of rest-mass 
energy to neutrinos is nearly maximal
(see Ruffert et al. 1997; Ruffert \& Janka 1999). 

Assuming that the average 
neutrino luminosity $\ave{L_\nu}$ at $t_{\rm sim}$ is representative for the
subsequent accretion phase, we obtain for $q_\nu$ numbers between 4 and 6\%
and total energies $E_\nu \sim \ave{L_\nu}t_{\rm acc}$ around 
$3\times 10^{52}\,$erg (Table~2). Annihilation of neutrino pairs, 
$\nu\bar\nu \to e^+e^-$, deposits energy at rates up to $\dot E_{\nu\bar\nu}\sim
2\times 10^{52}\,$erg/s in the vicinity of the black hole (Fig.~4). This
corresponds to total energies $E_{\nu\bar\nu}\sim \dot E_{\nu\bar\nu}t_{\rm acc}$
as high as $\sim 10^{51}\,$erg and annihilation efficiencies
$q_{\nu\bar\nu} = \dot E_{\nu\bar\nu}/\ave{L_\nu}$ of 1--3\%.  
These estimates should not change much if the different 
effects of general relativity on $\nu\bar\nu$ annihilation are taken into
account in combination (Ruffert \& Janka 1999; Asano \& Fukuyama 1999),
but general relativistic simulations of the merging are very important.
More energy could be pumped 
into the $e^\pm\gamma$ fireball when the black hole rotates rapidly
(Popham, Woosley, \& Fryer 1999) or if magnetic fields are able to tap the 
rotational energy of the accretion torus and of the black hole with 
higher efficiency than $\nu\bar\nu$ annihilation does (Blandford \& Znajek 1977). 
This seems to be necessary for the long and very energetic GRBs
(M\'esz\'aros, Rees, \& Wijers 1999; Brown et al. 1999; Lee, Wijers, \& 
Brown 1999).

\acknowledgments 
HTJ was supported by DFG grant SFB 375 f\"ur Astro-Teilchenphysik,
MR by a PPARC Advanced Fellowship, and
CLF by NASA (NAG5-8128) and the US DOE ASCI Program (W-7405-ENG-48).


\newpage

{\footnotesize
\begin{table*}[t]
\tabcolsep 2.5pt
\begin{center}
\centerline {\small TABLE~1}
\centerline{\footnotesize GRAVITY WAVES AND NEUTRINOS FROM NS/NS AND BH/NS MERGING}
\begin{tabular}{cccccccccccccccc}
\tableline\tableline\\[-3mm]
Model & Type & Masses & Spin & $t_{\rm sim}$ & $L_{\rm GW}^{\rm max}$  & $rh^{\rm max}$ & $E_{\rm GW}$ &
$L_{\nu_e}^{\rm max(av)}$ & $L_{\bar\nu_e}^{\rm max(av)}$ & $L_{\Sigma\nu_x}^{\rm max(av)}$ & $E_{\nu}$ & 
$kT^{\rm max}$ & $\ave{\epsilon_{\nu_e}}$ & $\ave{\epsilon_{\bar\nu_e}}$ & $\ave{\epsilon_{\nu_x}}$\\[1mm]
   &   & $(M_{\odot})$ &  & (ms) & $(10^4\,{{\rm foe}\over{\rm s}})^a$ & $(10^4{\rm cm})$ & (foe) &
$(100\,{{\rm foe}\over{\rm s}})$ & $(100\,{{\rm foe}\over{\rm s}})$ & $(100\,{{\rm foe}\over{\rm s}})$ &
(foe) & (MeV) & (MeV) & (MeV) & (MeV)\\[1mm]
\tableline\\[-3mm]
   S64   & NS/NS & 1.2+1.2  & solid & 10 & 0.7 & 5.5  & 14  & 0.3(0.2) & 0.9(0.5) & 0.3(0.2) & 0.8 & 35 & 12 & 18 & 26\\
   D64   & NS/NS & 1.2+1.8  & solid & 13 & 0.4 & 5.5  & 13  & 0.5(0.3) & 1.3(0.8) & 0.7(0.4) & 1.1 & 35 & 13 & 19 & 27\\
   V64   & NS/NS & 1.6+1.6  & anti  & 10 & 1.2 & 6.0  & 23  & 1.1(0.5) & 2.6(1.3) & 0.7(0.3) & 1.9 & 69 & 13 & 19 & 27\\
   A64   & NS/NS & 1.6+1.6  & none  & 10 & 2.1 & 8.6  & 52  & 0.9(0.5) & 2.6(1.3) & 1.4(0.6) & 2.3 & 39 & 12 & 18 & 26\\
   B64   & NS/NS & 1.6+1.6  & solid & 10 & 2.1 & 8.9  & 37  & 0.6(0.4) & 1.8(1.1) & 0.9(0.4) & 1.8 & 39 & 13 & 19 & 27\\
\tableline
   TN10  & BH/AD & 2.9+0.26 & solid & 15 & ... & ... & ...  & 0.5(0.4) & 1.3(0.9) & 0.6(0.2) & 0.8 & 15 &  9 & 13 & 21\\ 
\tableline
   C2.5  & BH/NS & 2.5+1.6  & anti  & 10 & 2.3 & 9.9  & 32  & 1.5(0.5) & 7.3(2.5) & 5.2(1.9) & 4.5 & 74 & 16 & 22 & 31\\
   A2.5  & BH/NS & 2.5+1.6  & none  & 10 & 2.0 & 9.9  & 50  & 1.8(0.5) & 6.4(2.2) & 3.1(1.3) & 3.6 & 65 & 15 & 22 & 31\\
   B2.5  & BH/NS & 2.5+1.6  & solid & 10 & 2.1 & 9.6  & 61  & 0.9(0.3) & 6.5(1.7) & 3.6(0.9) & 2.5 & 61 & 14 & 21 & 29\\
   C5    & BH/NS & 5.0+1.6  & anti  & 15 & 3.9 & 13.0 & 50  & 0.7(0.4) & 3.8(1.6) & 2.5(1.1) & 4.5 & 46 & 15 & 20 & 29\\
   A5    & BH/NS & 5.0+1.6  & none  & 20 & 3.2 & 14.8 & 102 & 0.7(0.2) & 4.4(1.5) & 2.8(0.8) & 4.5 & 51 & 16 & 24 & 31\\
   B5    & BH/NS & 5.0+1.6  & solid & 15 & 3.4 & 14.5 & 95  & 0.6(0.2) & 3.7(1.1) & 2.5(0.6) & 2.9 & 44 & 14 & 21 & 28\\
   C10   & BH/NS & 10.0+1.6 & anti  & 10 & 7.1 & 21.9 & 123 & 0.4(0.1) & 2.5(0.4) & 1.2(0.1) & 0.6 & 51 & 14 & 19 & 24\\ 
   A10   & BH/NS & 10.0+1.6 & none  & 10 & 6.9 & 26.2 & 168 & 0.2(0.1) & 2.5(0.5) & 1.2(0.2) & 0.7 & 50 & 14 & 20 & 26\\
   B10   & BH/NS & 10.0+1.6 & solid & 10 & 7.3 & 26.2 & 163 & 0.4(0.1) & 2.5(0.8) & 1.4(0.2) & 1.1 & 52 & 13 & 18 & 24\\[1mm]
\tableline
\end{tabular}
\end{center}
$^a$ 1 foe = $10^{51}\,$erg ({\bf f}ifty {\bf o}ne {\bf e}rg).
\end{table*}
}

\newpage

{\footnotesize
\begin{table*}[t]
\tabcolsep 3pt
\begin{center}
\centerline {\small TABLE~2}
\centerline{\footnotesize DISK FORMATION AND NEUTRINO ANNIHILATION}
\begin{tabular}{ccccccrrrcccccrrrr}
\tableline\tableline\\[-3mm]
Model & $t_{\rm ns}$ & $d_{\rm ns}$ & $M_{\rm ns}^{\rm min}$ & $\Delta M_{\rm ej}$ &
$M_{\rm d}$  & $\dot M_{\rm d}$ & $t_{\rm acc}$ & $\alpha_{\rm eff}$ &
$a_{\rm i}$ & $a_{\rm f}$ & $a_{\rm BH}^\infty$ &
$\ave{L_{\nu}}$ & $\dot E_{\nu\bar\nu}$ & $q_{\nu}$ & $q_{\nu\bar\nu}$ & 
$E_{\nu}$ & $E_{\nu\bar\nu}$\\
   & (ms) & (km) & $(M_{\odot})$ & $(M_{\odot}/100)$ &
   $(M_{\odot})$ & $(M_{\odot}/{\rm s})$ & (ms) & $(10^{-3})$ & 
&   &   & $(100\,{{\rm foe}\over{\rm s}})^a$ &
(foe/s) & (\%) & (\%) & 
 (foe) & (foe)\\[1mm]
\tableline\\[-3mm]
   S64   & 2.8  & 15 & ...  & 2.0    & ... & ... & ... & ...  & 0.98 & 0.75 & ...  & 1.5 & 1  & ...   & 1   & ...  & ...   \\
   D64   & 7.3  & 15 & ...  & 3.8    & ... & ... & ... & ...  & 0.87 & 0.69 & ...  & 2   & 2  & ...   & 1   & ...  & ...   \\
   V64   & 3.7  & 15 & ...  & 0.0085 & ... & ... & ... & ...  & 0.64 & 0.49 & ...  & 4   & 9  & ...   & 2   & ...  & ...   \\
   A64   & 1.7  & 15 & ...  & 0.23   & ... & ... & ... & ...  & 0.76 & 0.55 & ...  & 5   & 9  & ...   & 2   & ...  & ...   \\
   B64   & 1.6  & 15 & ...  & 2.4    & ... & ... & ... & ...  & 0.88 & 0.63 & ...  & 3   & 7  & ...   & 2   & ...  & ...   \\
\tableline
   TN10  & ...  & ...& ...  & ...    &0.26 & 5   & 53  & 4    & ...  & 0.42 & 0.59 & 1.2 &0.5 & 1.3   & 0.4 & 7    & 0.03  \\
\tableline
   C2.5  & 2.6  & 11 & 0.78 & 0.01   &0.26 & 6   & 43  & 4    & 0.65 & 0.47 & 0.60 &  7  & 20 & 6     & 3   & 30   &   0.9 \\
   A2.5  & 4.3  & 18 & 0.78 & 0.03   &0.33 &$<14$&$>24$& $<8$ & 0.67 & 0.39 & 0.56 &  7  & 20 & $>3$  & 3   & $>17$&$>0.5$ \\
   B2.5  & 6.0  & 23 & 0.78 & 0.2    &0.45 &$<35$&$>13$& $<14$& 0.69 & 0.38 & 0.61 &  7  & 20 & $>1$  & 3   & $>9$ &$>0.3$ \\
   C5    & 9.1  & 76 & 0.40 & 2.5    &0.38 & 5   & 76  & 5    & 0.44 & 0.27 & 0.42 &  4  & 8  & 4     & 2   & 30   &   0.6 \\
   A5    & 16.3 & 65 & 0.52 & 2.5    &0.49 & 6   & 82  & 4    & 0.45 & 0.17 & 0.37 &  4  & 8  & 4     & 2   & 33   &   0.7 \\
   B5    & 10.8 & 79 & 0.50 & 5.6    &0.45 & 6   & 75  & 5    & 0.46 & 0.19 & 0.38 &  4  & 8  & 4     & 2   & 30   &   0.6 \\
   C10   & 8.0  & 96 & 0.65 & 2.2    &0.67 &$<10$&$>67$& $<11$& 0.24 & 0.07 & 0.25 &  2  & 2  & $>1$  & 1   & $>13$&$>0.1$ \\
   A10   & 9.3  & 95 & 0.60 & 3.2    &0.56 &$<60$&$>9$ & $<82$& 0.25 & 0.07 & 0.22 &  2  & 2  & $>0.2$& 1   & $>2$ &$>0.02$\\
   B10   & 5.1  & 97 & 0.65 & 10.0   &0.47 & 3   & 160 & 5    & 0.25 & 0.11 & 0.23 &  2  & 2  & 4     & 1   & 32   &   0.3 \\[1mm]
\tableline
\end{tabular}
\end{center}
$^a$ 1 foe = $10^{51}\,$erg ({\bf f}ifty {\bf o}ne {\bf e}rg).
\end{table*}
}

\clearpage

\epsfxsize=15truecm \epsfbox{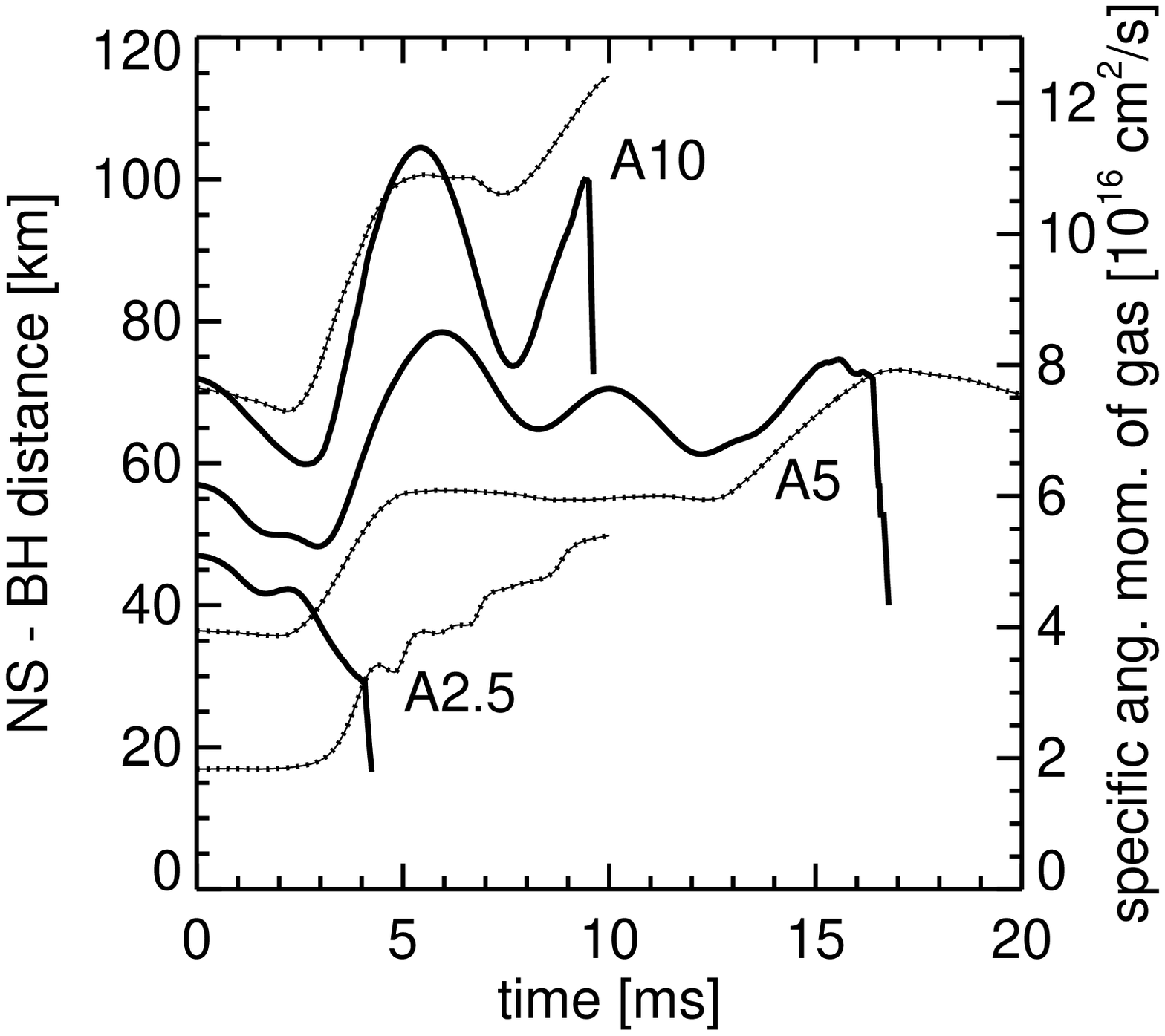}
\figcaption{Orbital separation between black hole and neutron star (solid lines)
and specific angular momentum of the gas on the grid (dotted lines) as
functions of time for Models A2.5, A5, and A10. The steep drop at the end of
the solid lines marks the moment when the neutron star is disrupted.}
\newpage

\epsfxsize=15truecm \epsfbox{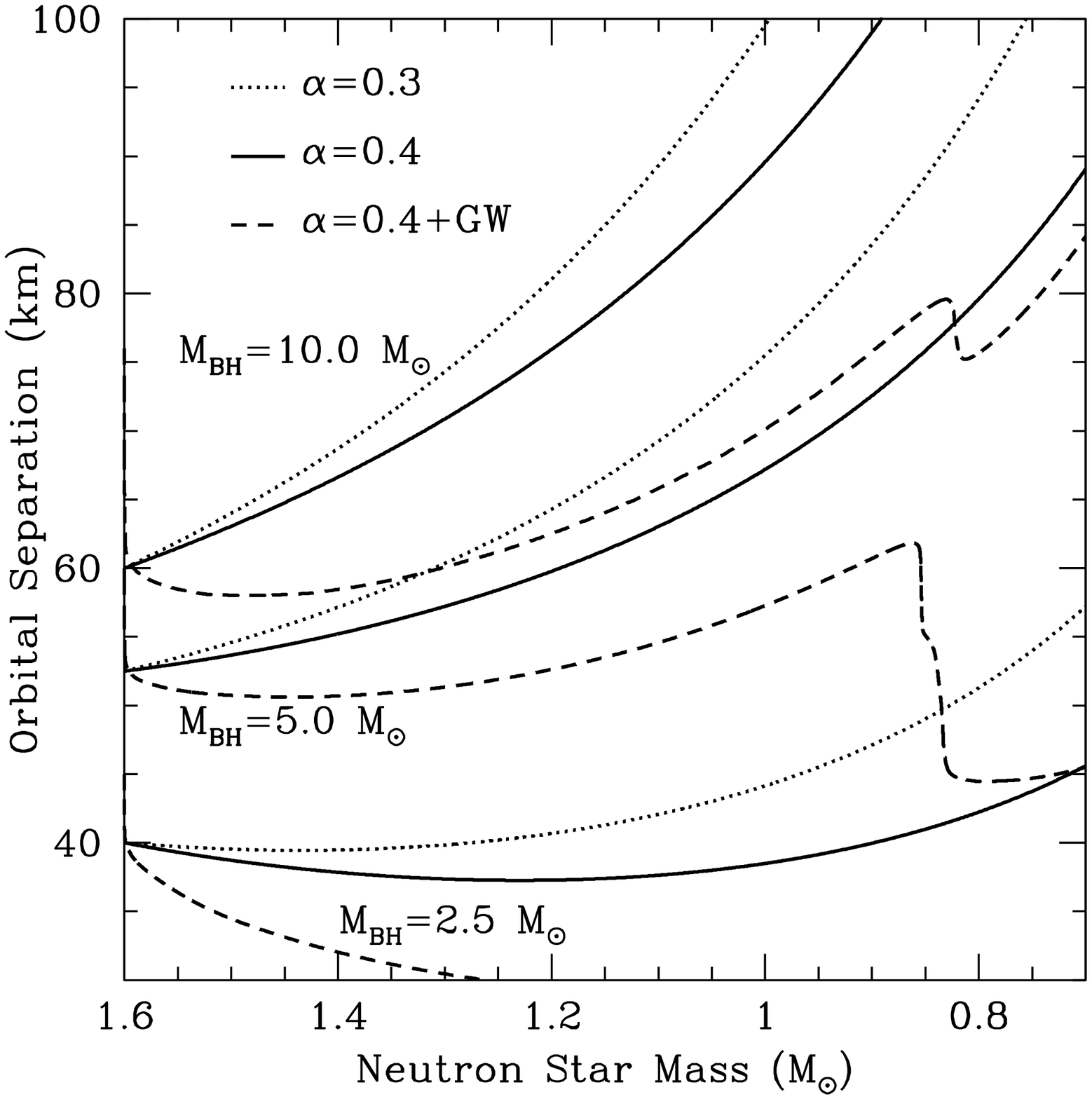}
\figcaption{Orbital separation as function of neutron star mass for different
initial black hole masses and values of parameter $\alpha$ in a simple analytic
model (see text). Note that
the total mass of the system, $M_{\rm BH}+M_{\rm NS}$ is constant along the lines.
Mass transfer leads to orbit widening only for $M_{\rm BH} = 5$ 
and $10\,M_{\odot}$, whereas GW emission decreases the separation. Combining
both effects (dashed lines) qualitatively explains the behavior shown 
for the hydro simulations in Fig.~1.}
\newpage

\epsfxsize=15truecm \epsfbox{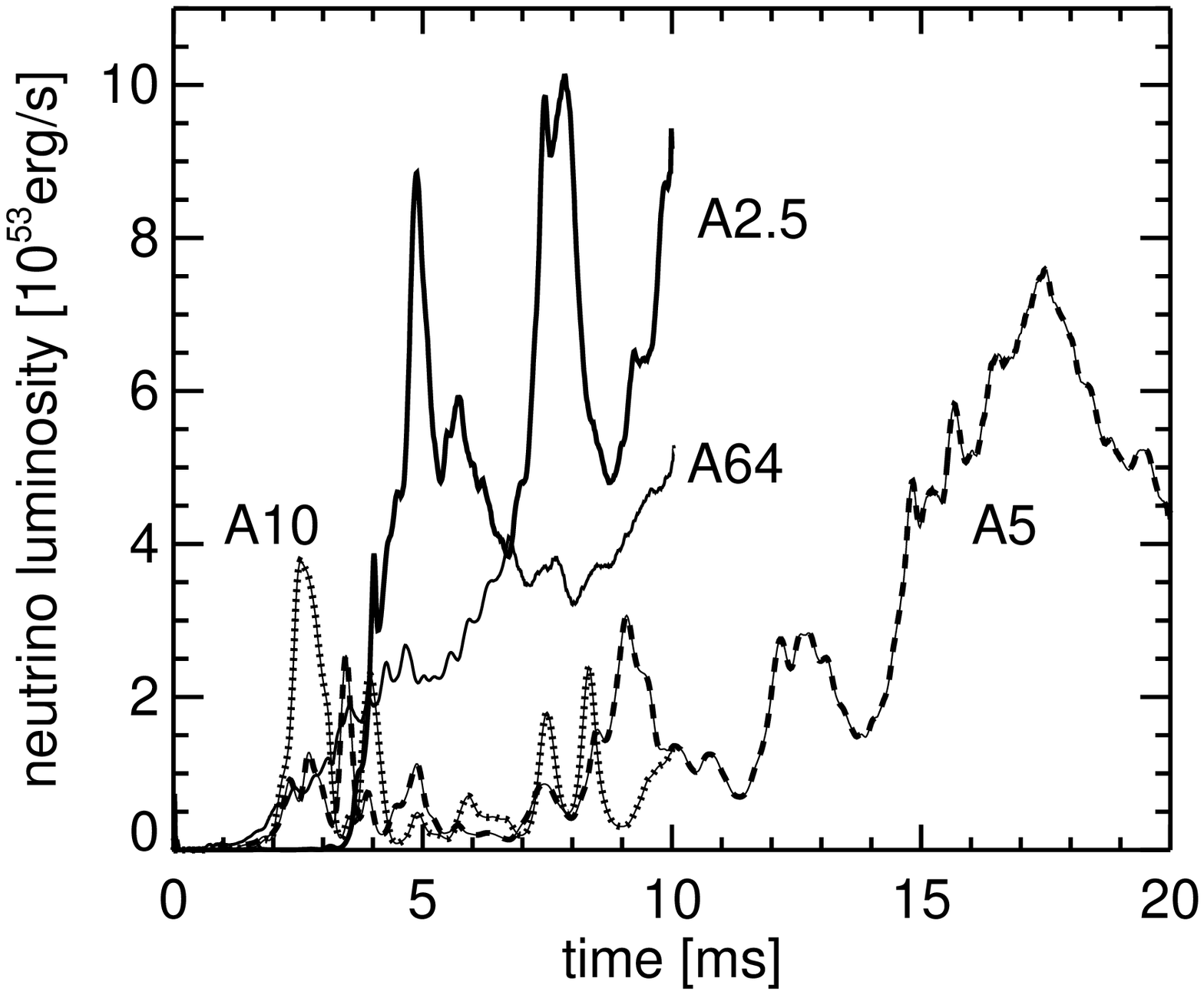}
\figcaption{Total neutrino luminosities as functions of time for BH/NS merger
Models A2.5, A5, and A10, and for the NS/NS merger Model A64.}
\newpage

\epsfxsize=15truecm \epsfbox{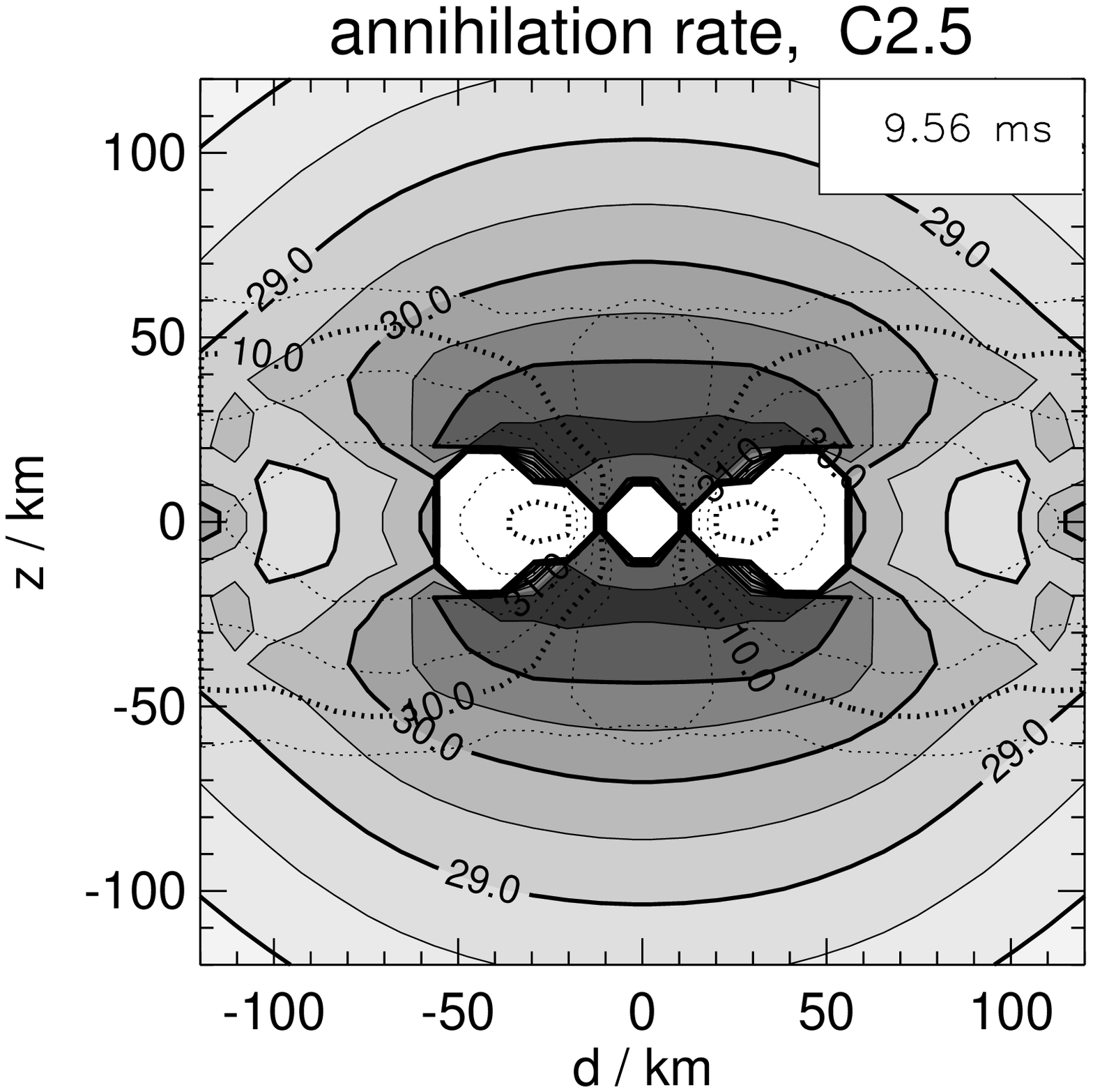}
\figcaption{Contours of the logarithm of the azimuthally averaged density distribution
(dotted lines) of the accretion torus around the black hole (indicated by the white
octagonal area at the center) and of the logarithm of the energy deposition rate per
cm$^3$ by $\nu\bar\nu$ annihilation into $e^+e^-$ pairs (solid lines) for the BH/NS
merger Model C2.5 at time 9.56$\,$ms. The contours are spaced in steps of 0.5~dex.
The integral energy deposition rate is $2\times 10^{52}\,$erg/s.}

\end{document}